# Anisotropic behaviour of S-wave and P-wave states of heavy quarkonia at finite magnetic field

Manohar Lal[a], Siddhartha Solanki[a], Rishabh Sharma[a], and Vineet Kumar Agotiya[a1]

[a]Department of Physics, Central University of Jharkhand Ranchi, India, 835 222

We studied the effect of momentum space anisotropy on heavy quarkonium states using an extended magnetized effective fugacity quasiparticle model (EQPM). Both the real and imaginary part of the potential has been modified through the dielectric function by including the anisotropic parameter $\xi$. The real part of the medium modified potential becomes more attractive in the presence of the anisotropy and constant magnetic field. The binding energy of the $1S$, $2S$, and $1P$ quarkonium states including anisotropy effects for both the oblate and the isotropic case were studied. We find that the binding energy of $Q\bar{Q}$ states becomes stronger in the presence of anisotropy. However, the magnetic field is found to reduce the binding energy. The thermal width of the charmonium and bottomonium $1S$ states have been studied at constant magnetic field $eB = 0.3 \ GeV^2$ for isotropic and prolate cases. The effect of magnetic field on the mass spectra of the $1P$ state for the oblate case was also examined. The dissociation temperature for the $1S$, $2S$, and $1P$ charmonium and bottomonium have been determined to be higher for the oblate case with respect to the isotropic case.

## I. INTRODUCTION

Quarkonium states are sensitive to several important features of the Quark Gluon Plasma (QGP), including Landau damping and energy loss as mentioned in ref. [1, 2, 3]. Based on the experimental observations, quarkonium suppression, among the other signatures, is regarded as the clear probe of the QGP [4, 5, 6, 7, 8]. In the five decades since the discovery of the $J/\psi$ in 1974 [9, 10], quarkonium dissociation due to color screening in the deconfined medium suggested by Matsui and Satz [11] has become a pioneering research area in the particle physics. Several studies such as [12, 13, 14, 15, 16, 17] discuss the important refinements essential to the study of quarkonium in the thermal Hot Quantum Chromodynamics (QCD) medium. The physics of the heavy quarkonia, a strongly interacting matter, in the last two decades has great interest in the presence of the magnetic field upto the scale square mass of the pion $m_\pi^2$ or even larger details of which can be found in ref. [18]. Such kind of the studies are very relevant to the highly dense astrophysical compact bodies like the magnetars [19] and also useful for the study of the cosmological aspects [20]. Besides of these facts, the main motivation to study the effect of magnetic field in the heavy quarkonia was triggered by the fact that the order of the magnitude of this field e.g. $eB = m_\pi^2$ at Relativistic Heavy Ion Collider (RHIC) and $eB = 15m_\pi^2$ at Large Hadron Collider(LHC) during the lead-lead collision could be produced in the laboratory when the two heavy ions, traveling nearly equal to the speed of light, colliding with each other at zero impact parameter in the colliding region [21, 22, 23, 24, 25]. It is believed that such a large magnitude of the magnetic field produces at very early stages of the universe shortly after the big bang. However, it is not certain how long these generated magnetic field and to what extent survives in the thermalisation process of the QGP formation. Various theoretical models-based studies as well as the Lattice Quantum Chromodynamics (LQCD) prediction, discuss several phenomena effecting the properties of the quarkonia in the presence of the background magnetic field. However, the effects related to the magnetic field are of particular interest because of the fact that the heavy quarkonia are very sensitive to the earlier condition of the universe. Studies such as [26, 27, 28, 29, 30, 31, 32, 33] briefly explained the quarkonium spectra and the production rates in the magnetic field regime. Since QCD with non-zero magnetic field $eB$ does not have a sign problem, one can obtain the QCD phase diagram in the $T - eB$ plane using Monte Carlo calculations of the LQCD as the first ab-initio principle. Since, we know that the magnetic field generally causes the momentum anisotropy which is responsible for the instability in the Yang-Mills fields. Hence, this momentum anisotropy has played a vital role in the evolution of the QCD medium. When the charged particle placed in the strong magnetic field, the energy associated with circular motion of the charged particle, because of Lorentz force, is discretized (quantized). These quantized energy levels, due to the magnetic field effect, are known as Landau levels. However, in the presence of the strong magnetic field $q_f eB >> T^2$ only lowest Landau levels are populated ($l = 0$). This indicates the importance of the lowest Landau level dynamics.

In the present case, we are working under strong magnetic field $q_f eB > T^2$ using the effective fugacity quasi-particle model to study the anisotropic behaviour of the heavy quarkonia. One of the remarkable reasons to include the momentum anisotropy is that the QGP produced in the non-central collision does not possess isotropy. Moreover, momentum anisotropy present at each and every stage of the heavy ion collisions. This fact triggered most to study the anisotropic effect on the quarkonium properties in magnetic field regime. Several authors [34, 35, 36, 37, 38, 39, 40, 41] studied various observables of the quark gluon plasma by considering the momentum anisotropy. Following the ref. [42, 43, 44], the anisotropy has been introduced at the levels of distribution function. The gluon self-energy used to obtain gluon propagator and in turn to determine the dielectric permittivity in the presence of the anisotropy. The present manuscript is organized in the following manner:

Quasi-particle Debye screening mass in the presence of magnetic field is briefly discussed in section II (A). The quark-antiquark potential in the anisotropic medium is described in section II (B). The effect of momentum space anisotropy on the binding energy, dissociation temperature, thermal width and the mass spectra of the quarkoinum states in the presence of magnetic field has been briefly studied in section III. We discuss the results of the present work in section IV. Finally, we conclude our work in section V.

## II. MODEL SETUP

### A. Effective Quasi-Particle Model Extension in magnetic field and Debye Screening Mass

In the quasi-particle description, the system of the interacting particles is supposed to be non-interacting or weakly interacting by means of the effective fugacity [45] or with the effective mass [46, 47]. Nambu-Jona-Laisino (NJL) and Polykov Nambu-Jona-Laisino (PNJL) quasi-particles models [48], self-consistent quasi-particles model [49, 50, 51] etc. include the effective masses. Here, we considered the effective fugacity quasi-particle model (EQPM), in the presence of magnetic field, which interprets the QCD EoS as non-interacting quasi-partons with effective fugacity parameter $z_g$ for gluons and $z_q$ for quarks encoding all the interacting effects taking place in the medium. The distribution function for quasi-gluons and the quasi-quarks/quasi-antiquarks [52] are given as:

$$f_{g/q} = \frac{f_{g/q}e^{-\beta E_p}}{1 \mp f_{g/q}e^{-\beta E_p}}. \tag{1}$$

It is noted that $E_p = |\vec{p}| = p$ for the gluons whereas

$$E_p = \sqrt{p^2 + m^2 + 2n|q_f eB|}$$

for quarks/antiquarks. Where '$l$' denotes the Landau levels. In high energy physics, $\hbar = c = K_B = 1$. Therefore, $\beta = T^{\perp}$. One can look the physical significance of the effective fugacity in the following dispersion relation:

$$\omega_g = T^2 \partial_T \ln(z_g) + p. \tag{2}$$

$$\omega_q = T^2 \partial_T \ln(z_g) + q P_z^2 + m^2 + 2l|q_f eB|. \tag{3}$$

The first term in the above equations Eq. 2 and Eq. 3 represents the collective excitation of the quasi-gluons and quasi-quarks(quasi-partons). Thus, it is inferred that the effective fugacity $z_g$ and $z_q$ describes the hot QCD medium effects. The effective fugacity like in the other quasi-particle model modify the $T^{\mu\nu}$ energy tensor as discussed in [53]. Now, the extended version of the Effective fugacity quasi-particles model in the presence of magnetic field requires the modification of the dispersion relation as defined in the above Eq. 2 and Eq. 3 by the relativistic discretized Landau Levels. Thus, in view of this, the quark/antiquark distribution function can be obtained as below:

$$f_q^o = \frac{z_q e^{-\beta \sqrt{p_z^2 + m^2 + 2n|q_f eB|}}}{1 + z_q e^{-\beta \sqrt{p_z^2 + m^2 + 2n|q_f eB|}}}. \tag{4}$$

The effect of the magnetic field $B = B z\hat{}$ is taken along the z-axis. Since the plasma contains both the charged and the quasi-neutral particles, hence it shows collective behaviour. The Debye mass is an important quantity to describe the screening of the color forces in the Hot QCD medium. The Debye screening mass can be defined as the ability of the plasma to shield out the electric potential applied to it. In the studies [54, 55, 56, 57], detailed definition of the Debye mass can be found. To determine the Debye mass, in terms of the magnetic field, we start from the gluon-self energy as below:

$$m_D^2 = \Pi_{00}(\omega = p, |\vec{p}| \to 0) \tag{5}$$

According to the [58], gluon self-energy modified as:

$$\Pi_{00}(\omega = p, |\vec{p}| \to 0) = \frac{g^2|eB|}{2\pi^2 T} \int_0^\infty dp_z f_q^0 (1 - f_q^0) \tag{6}$$

Thus, Debye mass for quarks using the distribution function defined by Eq. 4 is given below as:

$$m_D^2 = \frac{4\alpha}{\pi T}|eB| \int_0^\infty dp_z f_q^0 (1 - f_q^0). \tag{7}$$

Since the magnetic field has no effect on the gluons, therefore, the gluonic contribution to the Debye mass will remain unchanged. In other words, the distribution function for the gluons with or without magnetic field will remain intact. For perturbative QCD, in the presence of magnetic field, the Debye screening mass can be derived with the application of kinetic theory approach. Both these approaches provide the similar results for the Debye mass in the presence of the magnetic field. So, the Debye mass for the $n_f = 3$ and $N_c = 3$ will:

$$m_D^2 = 4\alpha\left(\frac{6T^2}{\pi}PolyLog[2, z_g] + \frac{3eB}{\pi}\frac{z_q}{1+z_q}\right) \qquad (8)$$

The Debye mass for the ideal EoS $[z_0 g = 1]$ representing non-interacting quarks and gluons becomes:

$$m_D^2 = 4\pi\alpha\left(T^2 + \frac{3eB}{2\pi^2}\right) \qquad (9)$$

$\alpha$ is two loop coupling constant depending upon the temperature [59] and is given below:

$$\alpha(T) = \frac{6\pi}{(33-2nf)ln(\frac{T}{\lambda_T})}\left\{1 - \frac{(3(153-19nf))}{(33-2nf)^2}\frac{ln\left(2ln\left(\frac{T}{\lambda_T}\right)\right)}{ln\left(\frac{T}{\lambda_T}\right)}\right\}$$

where $n_f$ denotes the number of flavor which is 3 in our case and $\lambda_T$ is the QCD renormalisation scale.

### B. Quark-antiquark potential in the anisotropic medium

The solution of the Scrodinger equation (SE), although a century has passed, is still an important tool for both the physicists and chemists. The SE played a vital role to obtained not only the energy spectrum of the di-atomic and poly-atomic molecule but also the spectrum of the heavy quarkonium system. The solution of the SE for the different potentials, as found in [60, 61, 62, 63], has been obtained using generalized Boopp's shift method and standard peturbation theory. In the present work, medium modified potential [64] has been used to investigate the properties of the heavy quarkonia. The Cornell potential having both Coulombic as well as the string part [64] is given by:

$$V(r) = -\frac{\alpha}{r} + \sigma r \qquad (10)$$

To modified this static potential (Eq.10), we use the Fourier transformation. In the above equation, $\alpha$ and $\sigma$ are the coupling constant and the string tension respectively. Here we take the two-loop coupling constant depending upon on the temperature. The value of the string constant has been taken $\sigma = 0.184\ GeV^2$. Whereas 'r' is the effective radius of the respective quarkonium states. The reason behind the modification of the potential is that the string tension doesn't vanish at or near the transition temperature $T_c$ and transition is just "crossover" from hadronic to quark-gluon plasma (QGP). Since the heavy ions collisions are non-central, the spatial anisoptropy generates at the very early stages. As the QGP expands or evolves with time, different pressure gradient arises which are responsible for mapping the spatial anisotropy to the momentum anisoptropy. In the present formalism, anisotropy has been introduced at the particle's distribution level. Following the studies [43, 44, 65], the isotropic function has been employed to determine the anisotropic distribution function given below:

$$f(\mathbf{p}) \to f_\xi(\mathbf{p}) = C_\xi\,f\left(\sqrt{p^2 + \xi(p\cdot\hat{n})^2}\right) \qquad (11)$$

where $f(\mathbf{p})$, denotes the effective fugacity quasi-particle isotropic distribution function [66, 67, 68], $\mathbf{\hat{n}}$ is unit vector representing the direction of momentum anisotropy. The parameter $\xi$ represents the anisotropy of the medium. For isotropic case $\xi = 0$, for oblate form and prolate form the parameter $\xi$ in the $\mathbf{\hat{n}}$ direction lies in the ranges $\xi > 0$ and $-1 < \xi < 0$ respectively. Since the effects of different equations of state (EoS) enter through the Debye mass ($m_D$). So, to make Debye mass intact from the effects of anisotropy, present in the medium, we use the following normalisation constant $C_\xi$. With this normalisation constant $C_\xi$, the Debye mass will remain same for both the isotropic and anisotropic cases [65]. Therefore, the normalization constant, $C_\xi$, is written as:

$$C_\xi = \begin{cases} \frac{\sqrt{|\xi|}}{tan^{-1}\sqrt{|\xi|}} & if\ -1 \le \xi < 0 \\ \frac{\sqrt{\xi}}{tan^{-1}\sqrt{\xi}} & if\ \xi \ge 0 \end{cases} \qquad (12)$$

Eq. 12 can be simplified as:

$$C_\xi = \frac{\sqrt{|\xi|}}{tan^{-1}\sqrt{|\xi|}}\quad for\ \xi \ge -1, \qquad (13)$$

For small anisotropic effect, $\xi$ can be written as:

$$C_\xi = \begin{cases} 1 - \xi + O(\xi^{\frac{3}{2}}) & if\ -1 \le \xi < 0 \\ 1 + \xi + O(\xi^{\frac{3}{2}}) & if\ \xi \ge 0 \end{cases} \qquad (14)$$

or simply in the small $\xi$ limit, we have:

$$C_\xi = 1 + \frac{\xi}{3} + O(\xi^2). \qquad (15)$$

Following the assumption given by [14, 69, 70], potential of the dissipative anisotropic medium has been modified in the Fourier transform by dividing it with medium dielectric permittivity, $\epsilon(\mathbf{k})$:

$$\hat{V}(k) = \frac{\bar{V}(k)}{\epsilon(k)}.$$ (16)

Taking the inverse of Fourier transform defined above, the in medium/corrected potential reads off:

$$V(r) = \int \frac{d^3 \mathbf{k}}{(2\pi)^{3/2}} (e^{i\mathbf{k}\cdot\mathbf{r}} - 1) \hat{V}(k).$$ (17)

$\bar{V}(k)$ is the Fourier transform of $V(r)$ defined by Eq. 10 and given as

$$\bar{V}(k) = -\sqrt{\frac{2}{\pi}} \left( \frac{\alpha}{k^2} + 2\frac{\sigma}{k^4} \right)$$ (18)

Now, from the temporal component of the gluon propagators [75], the dielectric tensor in the Fourier space can be written as below:

$$\epsilon^{-1}(\mathbf{k}) = -\lim_{\omega \to 0} k^2 \Delta^{00}(\omega, \mathbf{k}).$$ (20)

where $\Delta^{00}$ represents the static limit of the "00" component of the gluon propagators in the Coulomb gauge. Also Eq. 20, according to linear response theory, provides the relation between the dielectric permittivity and the $\Delta^{00}$. The real and the imaginary part of the dielectric tensor obtained from the real part of the retard propagator and imaginary part of symmetric propagators respectively [76], in the static limit, are given below by Eqs. 21 and 22:

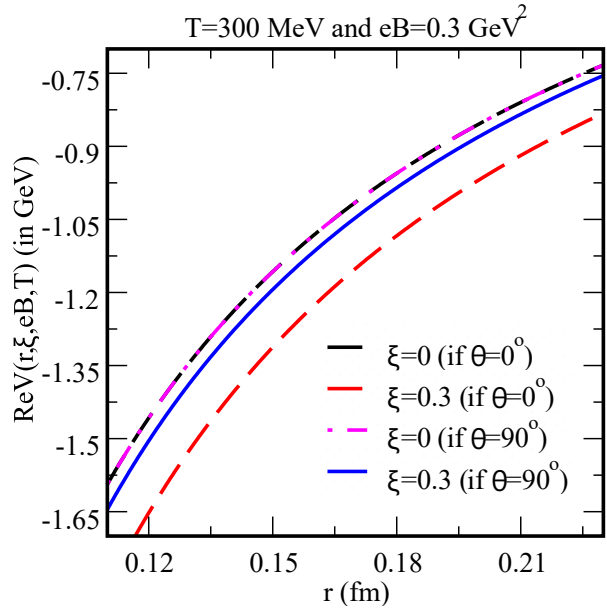

FIG. 1. Variation of real potential with distance 'r'(fm) for isotropic and the oblate case at fixed value of temperature and magnetic field

Thus, to obtained the modified form of the potential it is necessary to calculate the dielectric permittivity $\epsilon(\hat{k})$ and this can be done by two methods: (I) Using the gluon self-energy in finite temperature QCD [71, 72] and (II) Using Semi-classical transport theory (many particles kinetic theory up to one loop order) [44, 73, 74]. By using the above-mentioned methods, one can obtained the gluon self-energy $\Pi^{\mu\nu}$ which in turns provide static gluon propagator as given below:

$$\Delta^{\mu\nu} = k^2 g^{\mu\nu} - k^\mu k^\nu + \Pi^{\mu\nu}(\omega, \mathbf{k}).$$ (19)

$$\epsilon^{-1}(\mathbf{k}) = \frac{k^2}{k^2 + m_D^2} + k^2 \xi \left( \frac{1}{3(k^2 + m_D^2)} - \frac{m_D^2 (3\cos 2\theta_n - 1)}{6(k^2 + m_D^2)^2} \right).$$ (21)

$$\epsilon^{-1}(k) = \pi\, T\, m_D^2 \left( \frac{k^2}{k(k^2 + m_D^2)^2} - \xi k^2 \left( \frac{-1}{3k(k^2 + m_D^2)^2} + \frac{3\sin^2\theta_n}{4k(k^2 + m_D^2)^2} - \frac{2m_D^2(3\sin^2(\theta_n)-1)}{3k(k^2 + m_D^2)^3} \right) \right)$$ (22)

where

$$\cos(\theta_n) = \cos(\theta_r)\cos(\theta_{pr}) + \sin(\theta_r)\sin(\theta_{pr})\cos(\varphi_{pr}).$$ (23)

$\theta_n$ is the angle between the particle momentum **p**, and anisotropy direction **n̂**. $\theta_r$ is angle between **r**, and **n**. The azimuthal $\varphi_{pr}$ and the polar angle $\theta_{pr}$ lies between **p** and **r**. The term $m_D$ denotes the quasiparticle Debye mass which depends on the temperature and magnetic field and is briefly described in sec.III(A).

As the limit, $T \to 0$ real part of the potential goes to unity and when $\xi = 0$, the imaginary part becomes zero. With these limits, the modified form of the potential simply reduces to the Cornell potential. Now, by substituting the real part of the dielectric tensor $\epsilon^{-1}(\mathbf{k})$ defined by Eqs. 21 in Eq. 17, the real part of the interquark potential can be written as below:

$$Re[V(\mathbf{r}, \xi, T, eB)] = \left(1 + \frac{\xi}{3}\right)$$
$$\times \left(\frac{\sigma}{m_D} - \alpha\left(\frac{1}{s} + \frac{1}{2}\right)m_D\right) + \frac{\xi\,s}{16}\left(\frac{7}{3} - \cos(2\theta_r)\right), \tag{24}$$

After separating the coulombic ($\alpha$) and string ($\sigma$) term from the above equation, the real part of the potential will look like:

$$Re[V(\mathbf{r}, \xi, T, eB)] = \left(1 + \frac{\xi}{3}\right)\frac{\sigma}{m_D} - \alpha\left(\frac{1}{s} + \frac{1}{2}\right)m_D + \frac{\xi\,s}{16}\left(\frac{7}{3} - \cos(2\theta_r)\right) \tag{25}$$

Similarly, by putting the imaginary part of the dielectric tensor defined by Eqs. 22 in Eq. 17, the imaginary part of complex potential will become:

$$Im[V(r, \xi, T, eB)] = \pi T m_D^2 \int \frac{d^3\mathbf{k}}{(2\pi)^{3/2}} \left(e^{i\mathbf{k}\cdot\mathbf{r}} - 1\right)\left(-\sqrt{\frac{2}{\pi}}\frac{\alpha}{k^2} - \frac{4\sigma}{\sqrt{2\pi}k^4}\right)\left(\frac{k}{(k^2+m_D^2)^2} - \xi\left(\frac{-k}{3(k^2+m_D^2)^2} + \frac{3k\sin^2\theta_n}{4(k^2+m_D^2)^2} - \frac{2m_D^2 k(3\sin^2(\theta_n)-1)}{3(k^2+m_D^2)^3}\right)\right) \tag{26}$$

Again, separating the coulombic ($\alpha$) and string ($\sigma$) part of above equation 26, the imaginary potential can be rewritten as:

$$Im[V(r, \theta_r, T, eB)] = -\left(1 + \frac{\xi}{3}\right)T\left(\frac{\alpha s^2}{3} + \frac{\sigma s^4}{30m_D^2}\right)\log\left(\frac{1}{s}\right) + \xi T \log\left(\frac{1}{s}\right)\left\{\left(\frac{\alpha s^2}{10} + \frac{\sigma s^4}{140m_D^2}\right) - \cos^2\theta_r\left(\frac{\alpha s^2}{10} + \frac{\sigma s^4}{70m_D^2}\right)\right\} \tag{27}$$

## III. PROPERTIES OF HEAVY QUARKONIA

### A. Binding energy of the different quarkonium states

Real binding energies of the heavy quarkonium can be obtained by solving the Schrödinger equation with the first order perturbation in anisotropy parameter, $\xi$, as done in [75, 77, 78]. With this, the real binding energy ($E_B$) becomes:

$$Re[E_B(T, \xi, eB)] = \left(\frac{m_Q\sigma^2}{m_D^4 n^2} + \alpha m_D + \frac{\xi}{3}\left(\frac{m_Q\sigma^2}{m_D^4 n^2} + \alpha m_D + \frac{2m_Q\sigma^2}{m_D^4 n^2}\right)\right) \tag{28}$$

This is valid only for the ground and first excited states of the charmonium and bottomonium i.e. $J/\psi$, $\Upsilon$, $\Psi'$, and $\Upsilon'$. But in order to find the binding energies of $1P$ states of the charmonium and bottomonium, the correction term (potential and the kinetic energy) must be added to the binding energy of the $\Psi'$, and $\Upsilon'$. These correction term have been obtained by using the variational treatment method [79, 80, 81, 82, 83], in which the total energy consists of the kinetic energy correction and most importantly the correction added to the spin dependent potential which makes the $\psi$ and $\chi_c$ (both are first excited state) degenerate and hence obeys the Pauli's exclusion principle. The correction energy term, as found in the [81, 82, 84], is given below

TABLE I. Lower bound of dissociation temperature for isotropic case at $T_c = 197 MeV$

| Temperatures are in the unit of $T_c$ | | |
|---|---|---|
| For isotropic case ($\xi = 0$) | | |
| States | $eB = 0.3 GeV^2$ | $eB = 0.5 GeV^2$ | $eB = 0.7 GeV^2$ |
| $J/\psi$ | 1.5482 | 1.3578 | 1.0406 |
| $\Upsilon$ | 2.0304 | 1.8908 | 1.6751 |
| $\Psi'$ | 1.0532 | 0.7614 | 0.3553 |
| $\Upsilon'$ | 1.4340 | 1.2309 | 0.8756 |
| $\chi_c$ | 1.2944 | 1.0406 | 0.5203 |
| $\chi_b$ | 1.6116 | 1.4213 | 1.091 |

TABLE II. Lower bound of dissociation temperature for oblate case at $T_c = 197 MeV$

| Temperatures are in the unit of $T_c$ | | |
|---|---|---|
| For oblate case ($\xi = 0.3$) | | |
| States | $eB = 0.3 GeV^2$ | $eB = 0.5 GeV^2$ | $eB = 0.7 GeV^2$ |
| $J/\psi$ | 1.6497 | 1.4847 | 1.1928 |
| $\Upsilon$ | 2.1700 | 2.0431 | 1.8401 |
| $\Psi'$ | 1.1421 | 0.8756 | 0.4568 |
| $\Upsilon'$ | 1.5355 | 1.3451 | 1.0279 |
| $\chi_c$ | 1.3452 | 1.1040 | 0.6598 |
| $\chi_b$ | 1.6878 | 1.5101 | 1.2055 |

$$E_{\chi_c.\chi_b}^{Corr} = \frac{m_Q \sigma^2}{6m_D^3} \qquad (29)$$

Therefore, to evaluate the binding energy of the $1P$ states of heavy quarkonia, we add up this correction energy term to the binding energy of the $\psi^{'}, \Upsilon^{'}$ as defined in above equation. Hence, we have:

$$E_{(\chi_c.\chi_b)} = E_{(\Psi',\Upsilon')} + E_{(\chi_c.\chi_b)}^{corr} \qquad (30)$$

This implies

$$E_{(\chi_c.\chi_b)} = \left(\frac{m_Q \sigma^2}{m_D^4 n^2} + \alpha m_D + \frac{\xi}{3}\left(\frac{m_Q \sigma^2}{m_D^4 n^2} + \alpha m_D + \frac{2m_Q \sigma^2}{m_D^4 n^2}\right)\right) - \frac{m_Q \sigma^2}{6m_D^3} \qquad (31)$$

In the present work, the masses of $1P$ state as charmonia ($m_{\chi_c} = 1.865\,GeV$) and bottomonium ($m_{\chi_b} = 5.18\,GeV$) have been taken and details of which can found in [85] and references therein.

### B. *Dissociation of quarkonium states in in the presence of anisotropy and strong magnetic field*

Once we obtained the binding energies of the quarkonium states, it is customary to study the dissociation pattern of the quarkonium states when their binding energies becomes zero. But in the ref. [86], the authors have argued that it is not essential to have zero binding energy to dissociate the states but when the binding energy is less than the temperature ($E_B \leq T$), a state is weakly bound and hence, it is destroyed by the thermal fluctuations. The authors in [75, 86, 87], propose another condition for the dissociation of the quarkonium in [88] and reference therein, using thermal effects can be states and is $2E_B \leq \Gamma(T)$ or $\Gamma(T) \geq 2E_B$, where $\Gamma(T)$ is the thermal width of the respective state. Thus, there are two major criteria used to determine the dissociation temperature ($T_D$). The upper bound and the lower bound of the dissociation temperature, as one can found obtained by the following conditions:

$$E_{(J/\psi,\psi,\psi',\Upsilon')} = \left(\frac{m_Q \sigma^2}{m_D^4 n^2} + \alpha m_D + \frac{\xi}{3}\left(\frac{m_Q \sigma^2}{m_D^4 n^2} + \alpha m_D + \frac{2m_Q \sigma^2}{m_D^4 n^2}\right)\right)$$
$$= \begin{cases} T_D \to Upper\ bound\ of\ the\ quarkonium\ states \\ 3T_D \to Lower\ bound\ of the\ quarkonium\ states \end{cases} \qquad (32)$$

Eq. 32 is applicable for the $1S$ and $2S$ states of the heavy quarkonia. The dissociation temperature for the $1P$ states of charmonium and bottomonium has been calculated using the following relation:

$$E_{(\chi_c.\chi_b)} = \left(\frac{m_Q \sigma^2}{m_D^4 n^2} + \alpha m_D + \frac{\xi}{3}\left(\frac{m_Q \sigma^2}{m_D^4 n^2} + \alpha m_D + \frac{2m_Q \sigma^2}{m_D^4 n^2}\right)\right) + \frac{m_Q \sigma^2}{6m_D^3}$$
$$= \begin{cases} T_D \to Upper\ bound\ of\ the\ quarkonium\ states \\ 3T_D \to Lower\ bound\ of the\ quarkonium\ states \end{cases} \qquad (33)$$

### C. *Thermal width of $1S$ state of charmonium and bottomonium*

Since the imaginary part of the potential give rise to the thermal width which in turn also used to calculate the dissociation point of the quarkonium states. The thermal width of the quarkonium states can be obtained by using following ansatz:

$$\Gamma(T) = -\int d^3r \ |\Psi(r)|^2 \mathrm{Im}\ V(r) \qquad (34)$$

where, $\Psi(r)$ is the Coulombic wave function. The Coulombic wave function for ground state ($1S$, corresponding to $n = 1$ ($J/\psi$ and $\Upsilon$)) given as

$$\Psi_{1S}(r) = \frac{1}{\sqrt{\pi a_0^3}} e^{\frac{-r}{a_0}} \qquad (35)$$

where, $a_0 = 2/(m_Q \alpha)$ denotes Bohr radius of the quarkonia system. Now from Eq. 34, we have

$$\Gamma_{1S}(T) = \left(\frac{\xi}{3} + 1\right)\int d^3r |\Psi_{1S}(r)|^2 \left\{\alpha T s^2 \log\left(\frac{1}{s}\right)\left(\frac{1}{3} - \xi \frac{3-\cos 2\theta_r}{20}\right) + \frac{2\sigma T}{m_D^2} s^4 \log\left(\frac{1}{s}\right)\frac{1}{20}\left(\frac{1}{3} - \xi \frac{2-\cos 2\theta_r}{14}\right)\right\} \qquad (36)$$

Solving the above equation, we get the thermal width for $1S$-state as below:

$$\Gamma_{1S}(T) = T\left(\frac{\xi}{3} - 2\right)\frac{m_D^2}{\alpha m_Q^2}\Bigg\{\left(\frac{1}{6}\left(-25 + 12\gamma_E + 12\log(2) - 12\log(a_0) - 12\log(m_Q)\right) + \frac{3\sigma}{10\alpha^3 m_Q^2}\right) - (49 + 20\gamma_E - 12\log(a_0)) + 20\log\left(\frac{2}{m_Q}\right)\Bigg\} \qquad (37)$$

It is important to note that in ref. [75] while considering up to leading logarithmic order of imaginary potential, the authors too have taken the width up to the leading logarithmic. Thus, the dissociation width for $1S$-state would be of the form:

$$\Gamma_{1S}(T) = T\left(\frac{4}{\alpha m_Q^2} + \frac{12\sigma}{\alpha^4 m_Q^4}\right)\left(1 - \frac{\xi}{6}\right)m_D^2 \log\left(\frac{\alpha m_Q}{2m_D}\right) \qquad (38)$$

panel) at different values of the magnetic field.

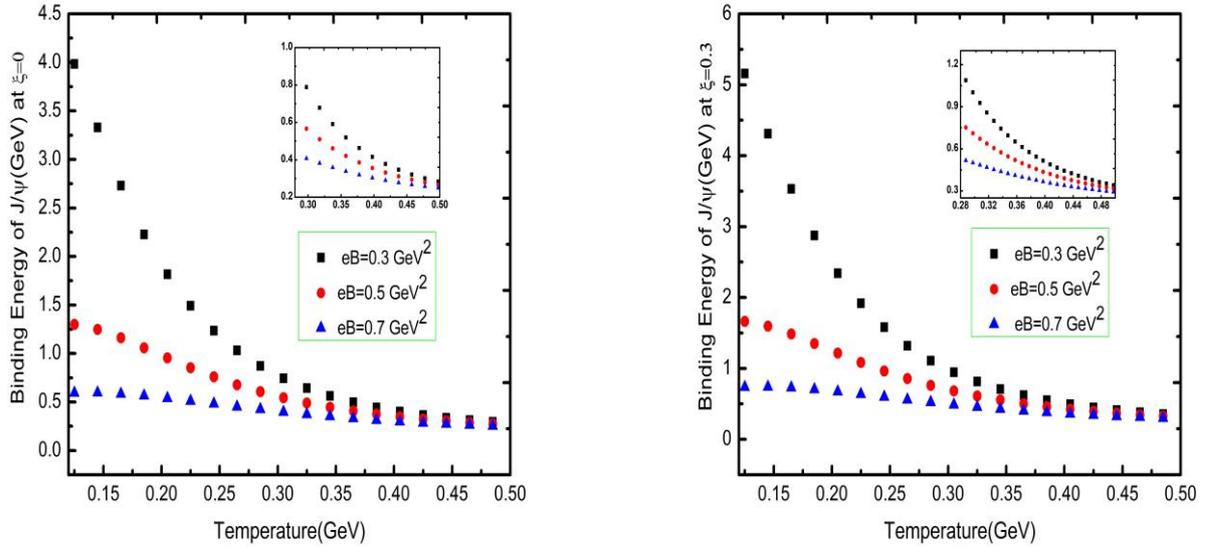

FIG. 2. Shows the variation of binding energy of the $J/\psi$ with temperature for the isotropic case(left panel) and oblate case(right panel) at different values of the magnetic field.

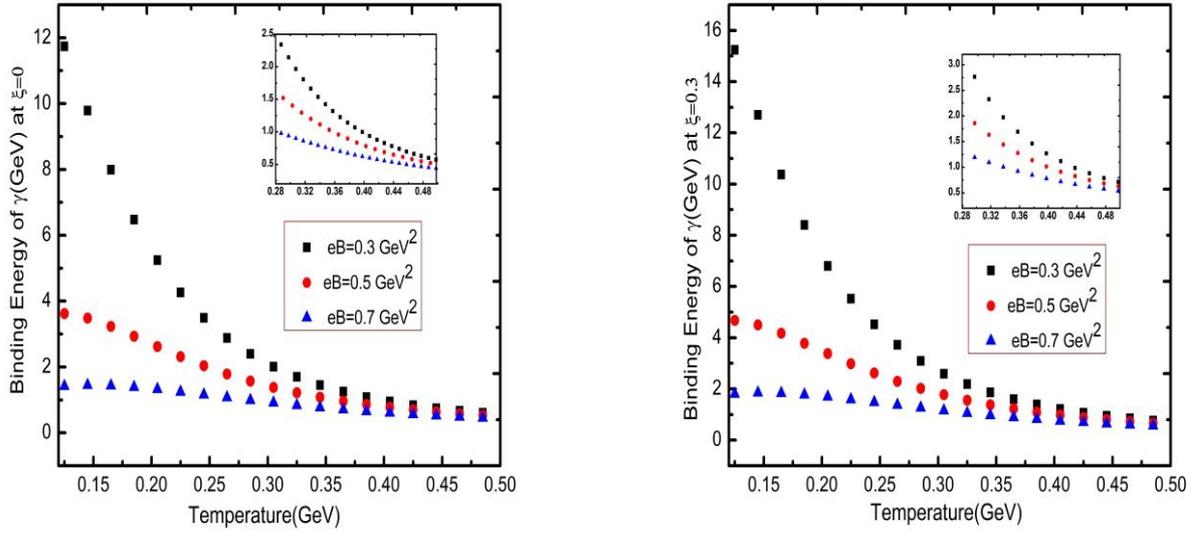

FIG. 3 Shows the variation of binding energy of the $\Upsilon$ with temperature for the isotropic case(left panel) and oblate case(right

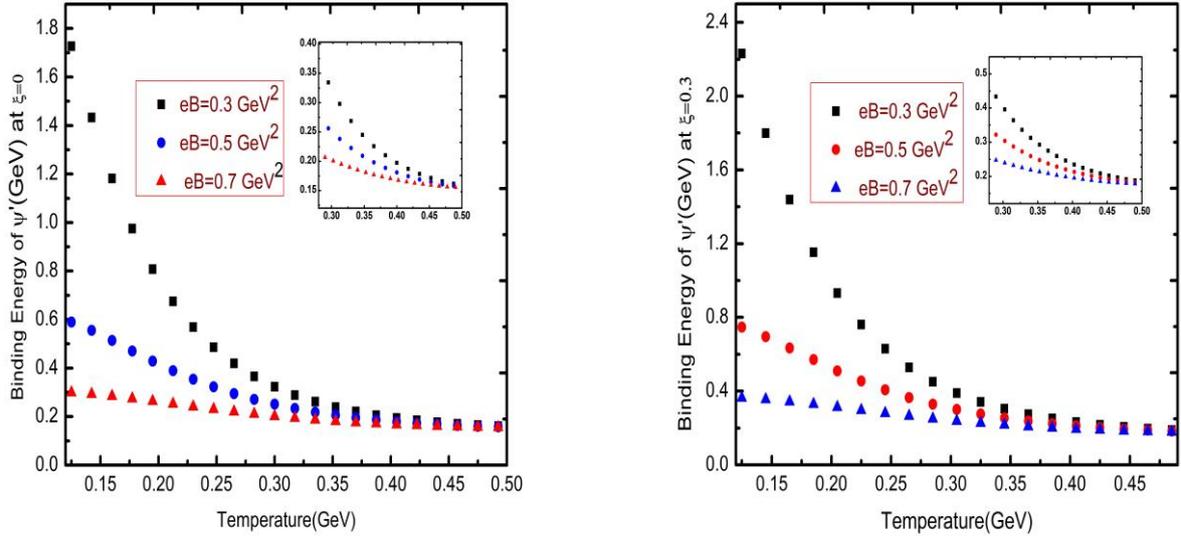

FIG. 4. Shows the variation of binding energy of the $\psi'$ with temperature for the isotropic case(left panel) and oblate case(right panel) at different values of the magnetic field.

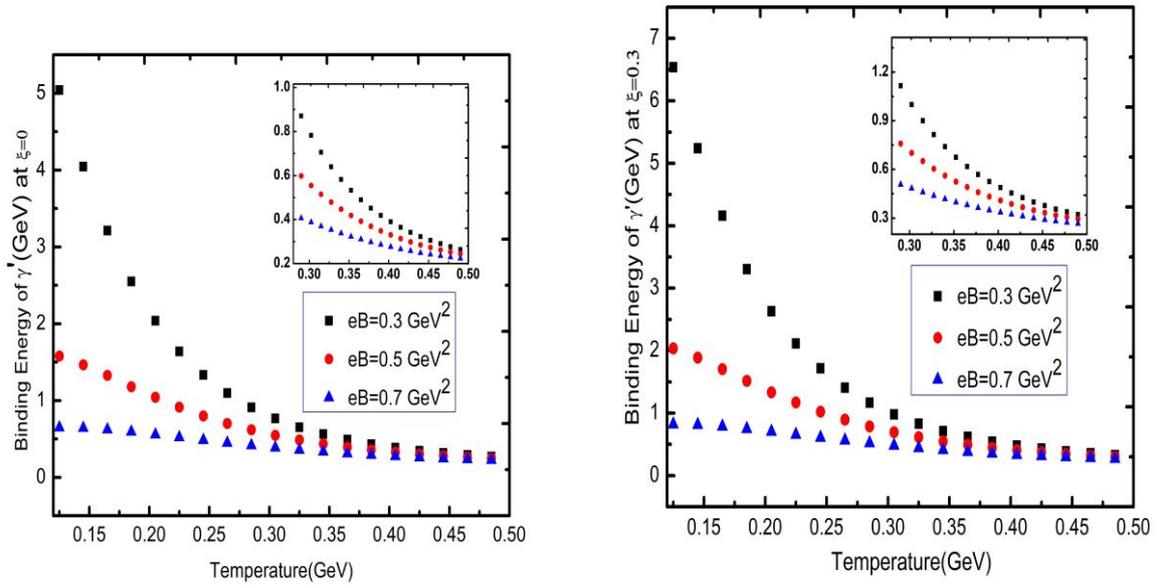

FIG. 5. Shows the variation of binding energy of the $\Upsilon'$ with temperature for the isotropic case(left panel) and oblate case(right panel) at different values of the magnetic field.

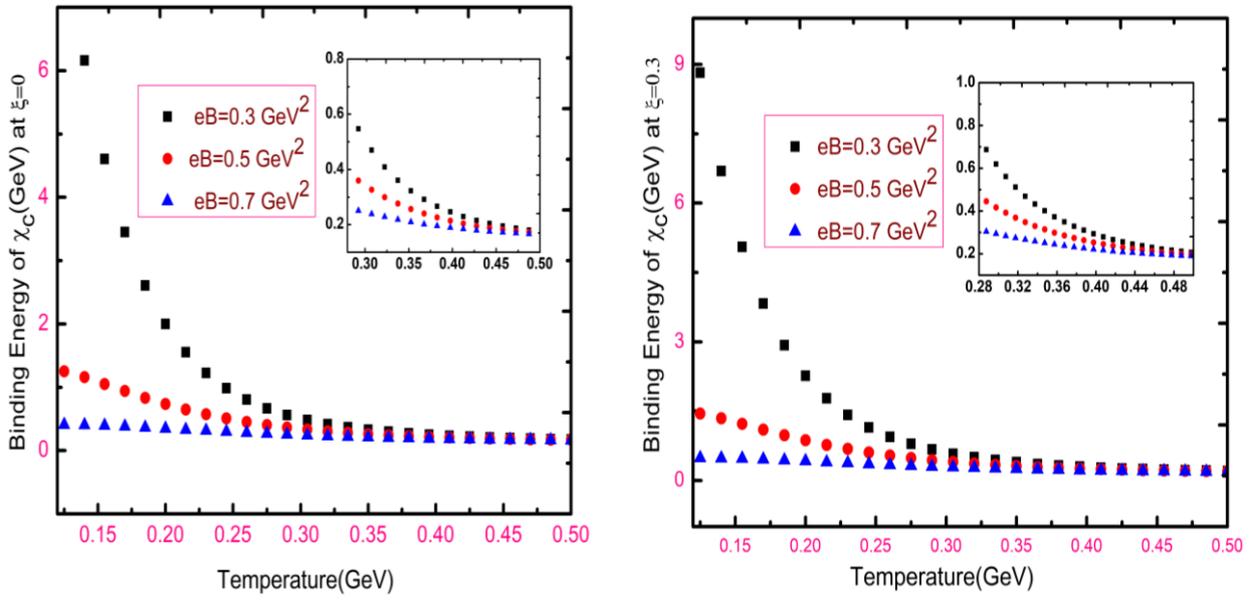

FIG. 6. Shows the variation of binding energy of the $\chi c$ with temperature for the isotropic case(left panel) and oblate case(right panel) at different values of the magnetic field.

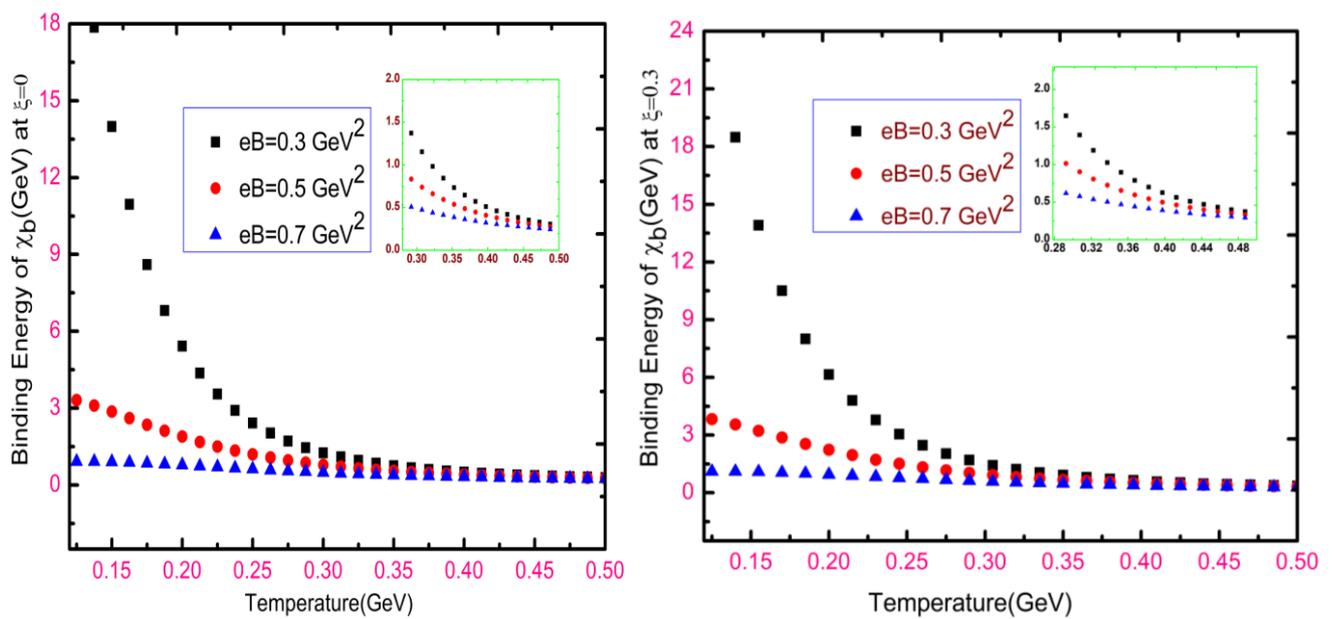

FIG. 7. Shows the variation of binding energy of the $\chi_b$ with temperature for the isotropic case (left panel) and oblate case (right panel) at different values of the magnetic field.

TABLE III. Upper bound of dissociation temperature for isotropic case at $T_c = 197 MeV$

| Temperatures are in the unit of $T_c$ | | |
|---|---|---|
| For isotropic case ($\xi = 0$) | | |
| States | $eB = 0.3 GeV^2$ | $eB = 0.5 GeV^2$ | $eB = 0.7 GeV^2$ |
| $J/\psi$ | 2.0304 | 1.8908 | 1.6757 |
| $\Upsilon$ | 2.6522 | 2.5507 | 2.3857 |
| $\Psi'$ | 1.4213 | 1.2182 | 0.9010 |
| $\Upsilon'$ | 1.8908 | 1.7385 | 1.5101 |
| $\chi_c$ | 1.6243 | 1.4467 | 1.1299 |
| $\chi_b$ | 2.0431 | 1.9035 | 1.6751 |

TABLE IV. Upper bound of dissociation temperature for oblate case at $T_c = 197 MeV$

| Temperatures are in the unit of $T_c$ | | |
|---|---|---|
| For oblate case ($\xi = 0.3$) | | |
| States | $eB = 0.3 GeV^2$ | $eB = 0.5 GeV^2$ | $eB = 0.7 GeV^2$ |
| $J/\psi$ | 2.1827 | 2.0558 | 1.8654 |
| $\Upsilon$ | 2.8299 | 2.7411 | 2.5888 |
| $\Psi'$ | 1.5355 | 1.3578 | 1.0659 |
| $\Upsilon'$ | 2.030 | 1.8908 | 1.6878 |
| $\chi_c$ | 1.7131 | 1.5482 | 1.2690 |
| $\chi_b$ | 2.1573 | 1.9035 | 1.6751 |

:

### D. Mass Spectra of the different quarkonium states in anisotropic hot QCD medium at finite magnetic field

The mass spectra of the different quarkonium states in the presence of magnetic field along with the effect of anisotropic parameter $\xi$ can be obtained by the relation:

$$M = 2m_Q + E_B. \qquad (39)$$

But in the current work we have calculated the mass spectra of only $1P$ state of the heavy quarkonia by using the following relation:

$$M = 2m_Q + E_{(\Psi', \Upsilon')} + E_{(\chi_c, \chi_b)}^{corr} \qquad (40)$$

Hence using Eq. 28 and 31 in Eq. 40, we have:

$$2m_Q + \left( \frac{m_Q \sigma^2}{m_D^4 n^2} + \alpha m_D + \frac{\xi}{3} \left( \frac{m_Q \sigma^2}{m_D^4 n^2} + \alpha m_D + \frac{2 m_Q \sigma^2}{m_D^4 n^2} \right) \right) + \frac{m_Q \sigma^2}{6 m_D^6} \qquad (41)$$

Where $m_Q$ ($m_{c,b}$) is the mass of the heavy quarkonia, $E_{(\Psi', \Upsilon')}$ is the binding energy of the $\psi'$, $\Upsilon'$ and $E_{(\chi_c, \chi_b)}^{corr}$ is the energy correction/ mass gap correction obtained using variational treatment method.

of the heavy quarkonia, $E$ ($\psi'$, $\Upsilon'$) is the binding energy of the $\psi'$, $\Upsilon'$ and $E_{\chi_c, \chi_b}^{Corr}$ is the energy correction/ mass gap correction obtained using variational treatment method.

### IV. RESULTS AND DISCUSSION

Heavy quarkonia properties have been investigated, by means of in medium modification to the Cornell potential (perturbative as well as non-perturbative), using extended quasi-particle approach in the presence of strong magnetic field limit $q_f eB >> T^2$. Since at the early stages of the Ultra-Relativistic Heavy ion Collisions (URHIC), the anisotropy arises in the beam direction as the system expands. At $\xi = 0$, the string term makes the real potential more attractive compared to the case when the potential is modified using Coulombic part only. This means that the respective quarkonium states becomes more bound with both the Coulombic and string part in comparison to the case when Coulombic part is modified alone. Here in the present work we have consider the weak anisotropy for the oblate case, $\xi = 0.3$ and isotropic case, $\xi = 0$ with the fixed value of the critical temperature $T_c = 197 MeV$ . The variation of the real potential with the distance (r in $fm$) has been shown in Figure 1 at $eB = 0.3$ $GeV^2$ and temperature $T = 300 MeV$ for oblate and isotropic case. For the isotropic case, $\xi = 0$, we have same variation for both parallel and perpendicular case. This is because of the fact that the system is expanding longitudinally. On the other hand, for the oblate case $\xi = 0.3$ the real potential have lower value for the parallel case($\theta = 0^o$) in comparison to the perpendicular case ($\theta = 90^o$). Figures 2, 3, 4, 5, 6 and 7 shows the variation of the binding energy of $J/\psi$, $\Upsilon$, $\psi'$ , $\Upsilon'$ , $\chi_c$ and $\chi_b$ with the temperature at finite values of the magnetic field for both the isotropic case $\xi = 0$ (left panel) and oblate case $\xi = 0.3$ (right panel) respectively. From all these figures, it has been deduced that the binding energy of all the quarkonium states $1S(J/\psi, \Upsilon)$, $2S$ ($\psi'$ , $\Upsilon'$ ) and $1P(\chi_c, \chi_b)$ decreases with the temperature. Also, as we goes from lower to higher magnetic field values, the binding energy also has lower values as can be seen from the Figures 2, 3, 4, 5, 6 and 7. However, it is noticed that binding energy of all the above mentioned states have higher values for the oblate case in competition to the isotropic case. In other words, the anisotropy seems to an additional handle to decipher the properties of the quarkonium states.

In anisotropic medium, the binding energy of the $Q\bar{Q}$ pair get stronger with increase in the anisotropy. This is due to the fact that the Debye screening mass in anisotropic medium is always much lower compared to the isotropic one. Hence, quarkonium states are strongly bound. Figure 8 shows the variation of the thermal width of charmonium($\Gamma_{J/\psi}$) (left panel) and bottomonium ($\Gamma_\Upsilon$) (right panel) at $eB = 0.3\ GeV\ ^2$ for the isotropic case and oblate case. It has been noticed from the Figure 8, that there is an increase in the thermal width with the temperature for both cases. Although thermal width has lower value for the oblate case in competition to the isotropic case. It is also noticed that the thermal width of the Upsilon ($\Gamma_\Upsilon$) is much smaller than the $J/\psi$ ($\Gamma_{J/\psi}$). This is due to the fact the bottomonium states are smaller in size and larger in masses than the charmonium states and hence get dissociated at higher temperatures. Mass spectra of the $\chi_c$ and $\chi_b$ has been shown in Figure 9. There is decreasing pattern of the mass spectra of $\chi_c$ and $\chi_b$ with the temperature at $\xi = 0.3$ at finite magnetic fields. Mass spectra of these states at $\xi = 0.3$ (oblate case) is found to be very closed to the particle data group 2018 [89]. The dissociation temperatures for the $1S(J/\psi,\Upsilon)$, $2S$ ($\psi'$, $\Upsilon'$) and $1P(\chi_c, \chi_b)$ has been given in the Table I, II, III, and IV. Lower bound of dissociation temperatures for different states have been shown in Table I and III for the isotropic case. Whereas the Table II and IV shows the different values of dissociation temperatures for the oblate case $\xi = 0.3$. In general, the dissociation temperatures decrease with the magnetic field. It is pertinent to mention here that for the oblate case dissociation temperatures have found to be higher in comparison to the isotropic case as seen from the tables.

## V. CONCLUSIONS

The dissociation process of the heavy quarkonium states $1S(J/\psi, \Upsilon)$, $2S$ ($\psi'$, $\Upsilon'$) and $1P(\chi_c, \chi_b)$ in the isotropic medium at finite magnetic field using extended quasi-particle model has been investigated. The real part of potential becomes more attractive in anistropic hot QCD medium compared to isotropic case at constant magnetic field. The binding energy of the different quarkonium states decreases with the temperature as well as with magnetic field for both the isotropic and oblate case. However, the binding energy has higher values for the oblate case $\xi = 0.3$ compared to isotropic case ($\xi = 0$). From the Figures 2, 3, 4, 5, 6 and 7 it is deduced that the binding energies of the $2S$ states ($\psi'$ ,$\Upsilon'$ ) are smaller than the binding energies of $1S(J/\psi, \Upsilon)$ and $1P(\chi_c, \chi_b)$ states for both the isotropic and oblate case in presence of magnetic field. The dissociation temperatures reduces as we increase the magnetic field for both the isotropic and oblate case. It is also found that the dissociation temperature for all the $1S$, $2S$ and $1P$ states have higher values for the oblate case $0 < \xi < 1$ compared to the case when $\xi = 0$. It is also noted here that the dissociation temperature of $1S$ and $1P$ states of charmonium and bottomonium are higher than the $2S$ states. This is because of the fact that the $\psi'$ and $\Upsilon'$ states are highly unstable or loosely bound states. The hierarchical order of dissociation temperatures ($T_D$) for the different quarkonium states is $T_D$'s ($1S$) > $T_D$'s ($1P$) > $T_D$'s ($2S$). Thermal width is found to increases with the temperature at constant magnetic field $eB = 0.3 GeV\ ^2$. The thermal width also increases with $\xi$ indicating quicker dissociation of the states. In future, this work will be extended to calculate the survival probability or the nuclear modification factor of different quarkonium states with respect to transverse momentum, centrality, and rapidity which is the key point to quantify the various properties of the medium produced during Heavy Ion Collisions (HIC) at LHC and RHIC.

## VI. DATA AVAILABILITY

The data used to support the findings of this study are available from the corresponding author upon request.

## VII. CONFLICTS OF INTEREST

The authors declare that they have no conflicts of interest regarding the publication of this paper.

## VIII. ACKNOWLEDGEMENT

VKA acknowledge the Science and Engineering research Board (SERB) Project No. **EEQ/2018/000181** New Delhi for providing the financial support. We record our sincere gratitude to the people of India for their generous support for the research in basic sciences

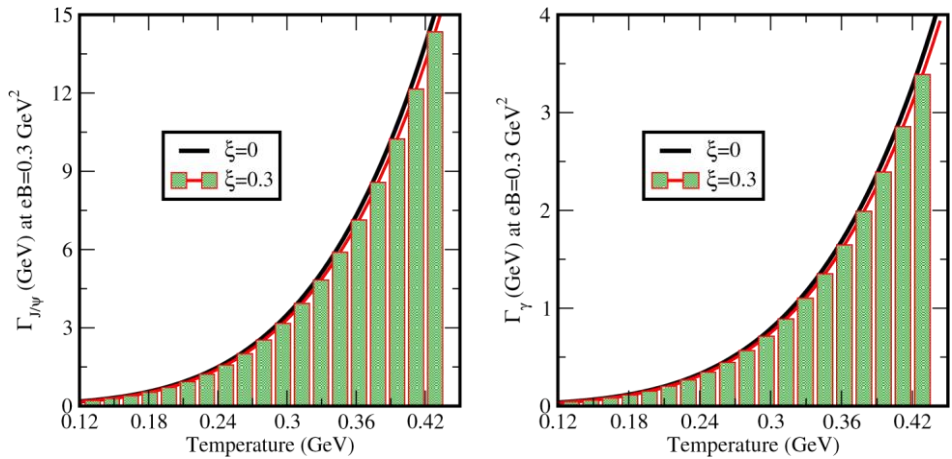

FIG. 8.   Variation of thermal width of $J/\psi$ and $\Upsilon$ with the temperature in the left and the right panel respectively at $\xi$ = 0.3 and $eB$ = $0.3 GeV^2$.

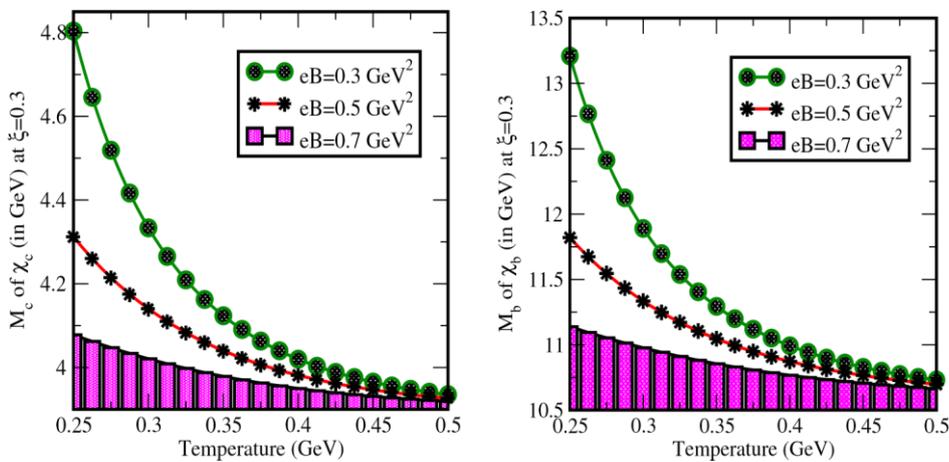

FIG. 9. Dependency of mass spectra of the $\chi_c$ and $\chi_b$ with temperature in the left and the right panel respectively at different magnetic field at $\xi$ = 0.3